\documentclass[mypaper,8pt,twoside]{CoAst}
\usepackage{epsf,graphicx,fancyhdr}
\usepackage{sfmath}
\renewcommand{\chaptermark}[1]{\markboth{#1}{}}

\newcommand{\kopf}{\small\itshape Comm. in Asteroseismology\\ Vol. 150, 2007}
\newcommand{\Authors}[1]{\begin{center}\normalsize\bf\sf #1 \end{center}}

\renewcommand{\author}[1]{\begin{center}\normalsize\bf\sf #1 \end{center}}
\newcommand{\Address}[1]{\begin{center}\small\sf #1 \end{center}}

\renewenvironment{abstract}{\section{Abstract}\normalsize\sf}{}
\newcommand{\References}[1]{\vspace{2.4mm}\begin{flushleft}{\large References\\}\vspace*{1mm}\small #1 \end{flushleft}}

\newcommand{\chapterDSSN}[2]{\chapter[\sf\normalsize #1\\ \footnotesize \hspace*{5mm}by #2 \sf\normalsize][]{#1\\}\rhead[\fancyplain{}{\sf\footnotesize \center{#1}}]{\fancyplain{}{\sffamily\thepage}}\lhead[\fancyplain{\kopf}{\sffamily\thepage}]{\fancyplain{\kopf}{\sf\footnotesize \center{#2}}}}

\newcommand{\figureDSSN}[5]{\begin{figure}[#4]
\centering
\includegraphics*[#5]{#1}
\caption{#2}
\label{#3}
\end{figure}}

\renewcommand{\chaptername}{}
\newcommand{\acknowledgments}[1]{\vspace*{5mm}\noindent\begin{bf}Acknowledgments. \end{bf} #1}

\def\rfr{\smallskip\par\noindent
        \hangindent=7truemm
        \hangafter=1}

\begin{document}
\sf

\chapterDSSN{Approaching asteroseismology of $\delta$ Scuti stars:
problems and prospects}{Jadwiga Daszy\'nska-Daszkiewicz}
\Authors{Jadwiga Daszy\'nska-Daszkiewicz} \Address{Instytut
Astronomiczny, Uniwersytet Wroc{\l}awski, ul. Kopernika 11, Poland}

\noindent
\begin{abstract}
The main obstacle in exploiting the frequency data of $\delta$ Sct
stars is difficulty in mode identification. The $\delta$ Sct
oscillation spectra, unlike those of the Sun or white dwarfs, do not
exhibit very regular patterns. Thus, the mode identification must
rely on sophisticated methods, which involve combined multi-passband
photometry and radial velocity data, with an unavoidable theoretical
input from stellar atmosphere models. Moreover, there are serious
uncertainties in theory of $\delta$ Sct stars that have to be
solved. Mode identification and determination of global and internal
structure parameters for $\delta$ Sct stars has to be done
simultaneously. I describe in some detail the methodology and
present  some recent results we obtained concerning degrees of
excited modes, global stellar parameters, and constraints on models
of subphotospheric convection, as well as effect of rotational mode
coupling.
\end{abstract}

\noindent
\section{Introduction}
$\delta$ Scuti stars are one of the most intensively studied group
of pulsating variables. In the HR diagram, they are located at the
intersection of the classical instability strip with the main
sequence, and somewhat above it. It was recognised many years ago
that pulsations of these objects, as other classical variables, are
driven by the $\kappa$-mechanism acting in the HeII ionization zone.
Excited are low-order p- and g-modes with periods ranging from 0.02
d to 0.3 d.

Over the last 20 years, many multisite observations of these stars
were carried out by DSN and WET networks. These campaigns have
resulted in a growing number of detected oscillation frequencies. On
the basis of these data several attempts were made towards
construction of asteroseismic models of certain multimodal
pulsators. One of such objects was XX Pyx, for which Pamyatnykh et
al. (1998) tried to construct seismic model without the $\ell$
identification from photometry or spectroscopy. Another example,
$\theta^2$ Tau, is a binary system consisting of an evolved
(primary) and main sequence A-type (secondary) stars (Breger et al.
2002), both inside the instability strip. The most multimodal and
most promising object for asteroseismology of $\delta$ Sct stars is
FG Vir. This star was studied by Guzik \& Bradley (1995), Viskum et
al. (1998), Breger et al. (1999) and Templeton et al. (2001). Recent
large photometric and spectroscopic campaigns, organised in the
years 2002-2004 by Breger et al. (2005) and Zima et al. (2006),
increased the number of independent oscillation frequencies of FG
Vir to 67. In spite of all these efforts we still do not have a good
seismic model for any $\delta$ Sct star. So far, not much has been
learnt from these rich oscillation spectra. There are still problems
with mode identification of excited modes as well as large
uncertainties in modelling $\delta$ Sct pulsation to exploit the
frequency data for constraining stellar model. The most important
aspects are: turbulent convection-pulsation interaction, effects of
rotation, mechanism of mode selection, diffusion.

In this paper, I outline the method which gives simultaneously mode
identification and constraints on stellar parameters and convection.
I discuss also effects of uncertainties arising from the atmospheric
models and, briefly, effects of rotational mode coupling on mode
identification.

\noindent
\section{Mode identification}
In the case of main sequence pulsators, the most widely used tools
for mode identification are pulsation amplitudes and phases derived
from observed variations in photometric passbands and in the radial
velocity. If effects of rotation can be neglected, the amplitude
ratio $vs.$ phase difference diagrams can lead to the $\ell$ degree
determination, and they are independent of the azimuthal order, $m$,
and the inclination angle. As was shown by Daszy\'nska-Daszkiewicz,
Dziembowski \& Pamyatnykh (2003) (Paper I), in the case of $\delta$
Sct variables the photometric amplitudes and phases are very
sensitive to the treatment of convection in the outer layers. This
is because in calculating these observables one has to make use of
the complex parameter, $f$, giving the ratio of the local flux
variation to the radial displacement at the photosphere. The $f$
parameter is obtained in the framework of linear nonadiabatic theory
of stellar oscillation and, in the case of $\delta$ Sct stars,
exhibits strong dependence on convection, as was already emphasised
by Balona \& Evers (1999). To avoid this problem, in Paper I we
invented a method of simultaneous determination of the $\ell$ degree
and $f$ parameter from multi-colour photometry and radial velocity
data. The procedure consists of solving the set of observational
equations for complex photometric amplitudes in a number of
passbands, $\lambda$, \figureDSSN{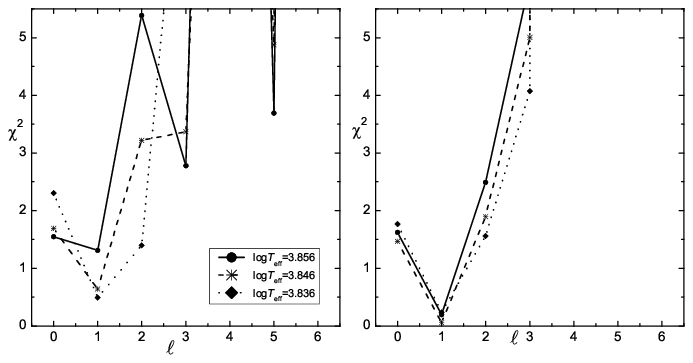}{ Values of
$\chi^2$ as a function of $\ell$ for the 9.897 d$^{-1}$ frequency
excited in $\beta$ Cas, for $M=1.95 M_{\odot}$ and three effective
temperatures. In the left panel, $\chi^2$ was calculated with the
Kurucz models and in the right panel with the NEMO2003 models.}
{fig1}{!ht}{clip,width=\textwidth}
\begin{equation}
{\cal D}_{\ell}^{\lambda} ({\tilde\varepsilon} f)
+{\cal E}_{\ell}^{\lambda} {\tilde\varepsilon} = A^{\lambda},
\end{equation}
where
$${\tilde\varepsilon} = \varepsilon Y^m_{\ell}(i,0),$$
$$ {\cal D}_{\ell}^{\lambda} = \frac14 b_{\ell}^{\lambda}
\frac{\partial \log ( {\cal F}_\lambda |b_{\ell}^{\lambda}| ) }
{\partial\log T_{\rm{eff}}}$$
$$ {\cal E}_{\ell}^{\lambda} =  b_{\ell}^{\lambda}
\left[ (2+\ell )(1-\ell ) - \left( \frac{\omega^2 R^3}{G M} + 2 \right)
\frac{\partial \log ( {\cal F}_\lambda
|b_{\ell}^{\lambda}| ) }{\partial\log g} \right]. $$
Derivatives of the monochromatic flux, ${\cal F}_\lambda(T_{\rm
eff},\log g)$, are calculated from static atmosphere models (Kurucz,
NEMO2003, Phoenix). In general, they depend also on the metallicity
parameter [m/H] and microturbulence velocity $\xi_t$. If the
spectroscopic data exist, the above set of equations can be
supplemented with the expression for the radial velocity (the first
moment of line profile, ${\cal M}_1^{\lambda}$),
\begin{equation}
{\rm i}\omega R \left( u_{\ell}^{\lambda}
+ \frac{GM}{R^3\omega^2} v_{\ell}^{\lambda} \right)
{\tilde\varepsilon}={\cal M}_1^{\lambda}.
\end{equation}
In the above expressions, $\varepsilon$ is the intrinsic mode
amplitude, $i$ is the inclination angle and
$b_{\ell}^{\lambda},~u_{\ell}^{\lambda},~v_{\ell}^{\lambda}$ are
disc-averaging factors weighted by the limb-darkening
$h_\lambda(T_{\rm eff},\log g)$. For the limb-darkening law we use
the Claret nonlinear formula. Each passband, $\lambda$, yields
r.h.s. of Eqs (1). The radial velocity data yield r.h.s. of Eq. (2).
Then, the system is solved by the least square method assuming trial
values of $\ell$. The $\ell$ identification is based on
$\chi^2(\ell)$ minimisation and quantities to be determined are:
$\tilde\varepsilon$ and $(\tilde\varepsilon f)$. In Paper I we
applied our method to three $\delta$ Sct stars: $\beta$ Cas, 20 CVn
and AB Cas, for their dominant frequencies. To this end we used
amplitudes and phases in four Str\"omgren passbands. In all cases
the identification of $\ell$ was unique. As an example, in Fig. 1 we
plot the $\chi^2(\ell)$ dependence for one frequency observed in the
star $\beta$ Cas. In the two panels, the effect of using atmospheric
models from different sources is shown. In the left panel, the
$\chi^2(\ell)$ was obtained adopting the Kurucz models, whereas in
the right one, adopting the Vienna models (NEMO2003). The method
works also in the case of multiperiodic pulsators. In
Daszy\'nska-Daszkiewicz et al. (2005) (Paper II) we applied the
method to the most multiperiodic $\delta$ Sct star FG Vir. Combining
the $vy$ Str\"omgren photometry and radial velocity data for twelve
modes, we arrived at a unique identification of $\ell$ in six cases,
and we obtained the constraint $\ell\le 2$ in other six.

The most important property of our method is that the identification
of the spherical harmonic degree, $\ell$, is independent of any
input from nonadiabatic pulsation calculations. Moreover, the method
uses simultaneously photometry and spectroscopy by combining these
data into the system of observational equations. For more details we
refer readers to Papers I and II.

\noindent
\section{Constraints on convection}
The above outlined method constitutes also a way of inferring $f$
from observations. The value of $f$, describing bolometric flux
perturbation, is determined in the pulsation driving zone, where the
thermal time scale is comparable with the pulsation period. It means
that this parameter is sensitive to properties of subphotospheric
layers which are poorly probed by oscillation frequencies. In
general, the $f$ parameter depends on: mean stellar parameters,
chemical composition, stellar convection and opacities. Thus, the
strong sensitivity of the $f$ parameter on convection in the case of
$\delta$ Sct pulsators may be considered as an advantage. Once we
know the empirical $f$ values, we can compare them with their
theoretical counterparts, and obtain valuable constraints on
convection in subphotospheric layers.

In Paper I, we succeeded in extracting the $f$ parameter photometric
observations for all studied $\delta$ Sct stars: $\beta$ Cas, 20 CVn
and AB Cas. We adopted Kurucz models of stellar atmospheres. The
pulsation calculations were made assuming a simplistic approach: the
mixing-length theory and the convective flux freezing approximation.
In comparison of empirical $f$ values with the theoretical ones
calculated with various values of the MLT parameter $\alpha$, we met
with a problem in reproducing both the real and imaginary part of
$f$ with the same value of $\alpha$. The general result was that
observed values of $f_R$ were close to those calculated with
$\alpha=0$, whereas values $f_I$ preferred rather higher values of
$\alpha$. The disagreement appeared to be mostly correlated with the
uncertainties in the atmospheric models, which I discuss in the next
section.

In Paper II we applied the method of simultaneous extracting $\ell$
and $f$ from observations for the most multiperiodic $\delta$ Sct
star FG Vir. We relied on NEMO2003 atmosphere models. Combining $vy$
Str\"omgren photometry and radial velocity data, we extracted the
$f$ parameter for twelve frequencies and compared them with the
theoretical values calculated assuming two different treatments of
convection. The first one was the standard mixing-length theory and
the convective flux freezing approximation, as in Paper I. As the
second one, we considered a non-local time-dependent generalisation
of MLT by Gough (1977). In the first case the agreement was found
for models with $\alpha\approx0.0$, which is an evidence that
convection in the outer layers of FG Vir is relatively inefficient.
In the second case, which includes convection dynamics, the
agreement was possible also with larger values of $\alpha$, but
smaller ones ($\alpha\le0.5$) were still favoured as can be seen
from Fig.2 (taken from Paper II).
\figureDSSN{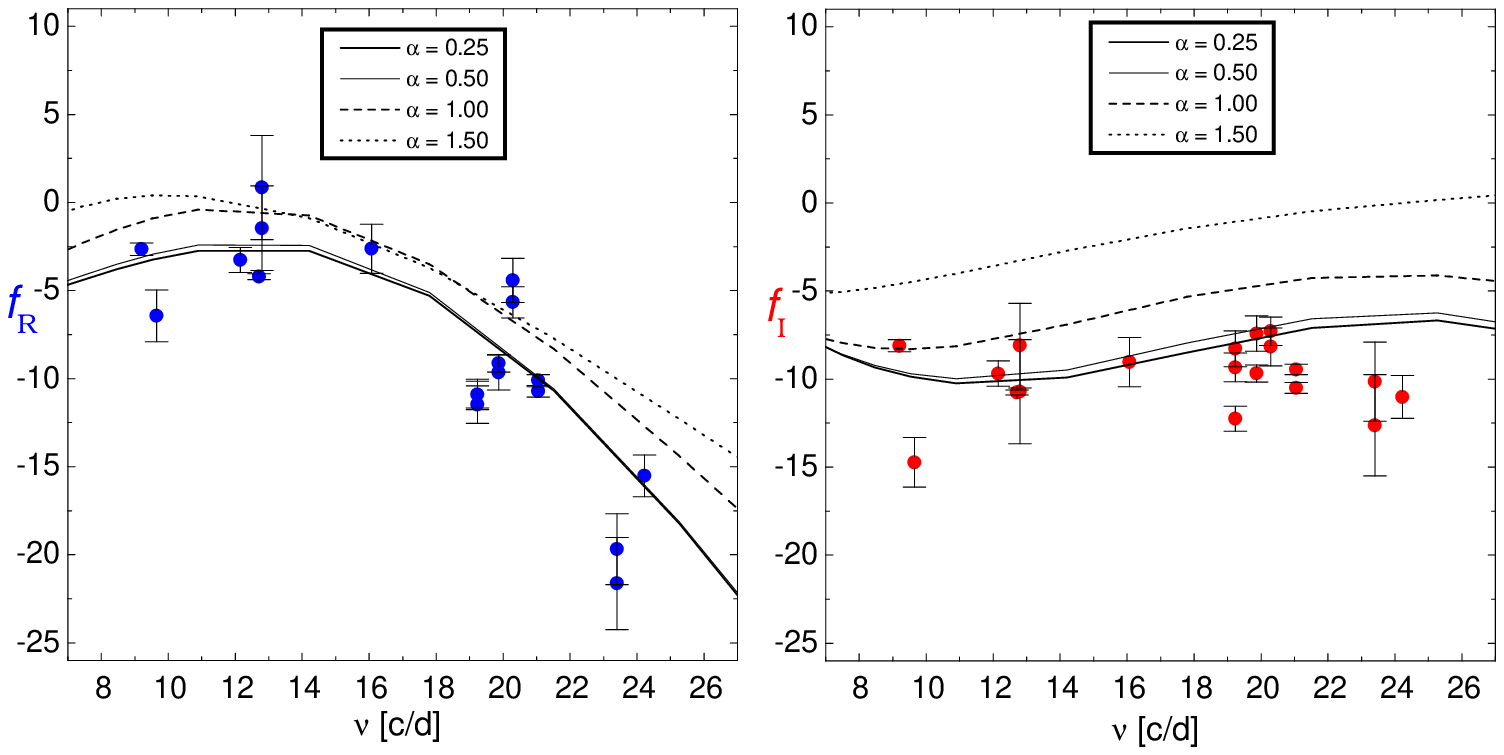}{ The empirical $f$ values (dots with
error bars) and the theoretical ones calculated for four values of
the MLT parameter $\alpha$, adopting a nonlocal, time-dependent
formulation of MLT. The real and imaginary parts of $f$ are shown in
the left and the right panels, respectively.
}{fig2}{!ht}{clip,height=5.3cm,width=\textwidth}

Modelling of $\delta$ Sct type pulsation with time-dependent
convection treatment can be found also in several other papers, e.g.
Grigah\`cene et al. 2005, Dupret et al. 2005a, Dupret et al. 2005b.

\noindent
\section{Uncertainties from atmospheric models}
To calculate pulsation amplitudes and phases of photometric and
radial velocity variations, one needs an input from atmospheric
models. As mentioned in the previous section, these are
monochromatic flux derivatives over effective temperature,
$\alpha_T$, and gravity, $\alpha_g$, as well as the limb-darkening
law, $h_\lambda$. In Fig. 3 we can see how non-smooth derivatives
$\alpha_T$, calculated from Kurucz models (left panel) can produce
artificial minima of $\chi^2$ derived from our method for a dominant
mode of FG Vir. Derivatives obtained from NEMO2003 models (right
panel) are smooth and only one $\chi^2$ minimum appears. In this
case, we show also the effect of microturbulence velocity, $\xi_t$,
on the location of minimum of $\chi^2(T_{\rm eff})$. The non-smooth
flux derivatives affect also the inferred  values of $f$. This is
illustrated in Fig. 4 where empirical $f$ values for $\beta$ Cas
obtained using Kurucz and Vienna models are compared with
theoretical ones calculated for five values of the MLT parameter,
$\alpha$. We can see that in the case of Vienna models both real and
imaginary parts of $f$ are reproduced with the models assuming
inefficient convection ($\alpha\approx0.0$)
\figureDSSN{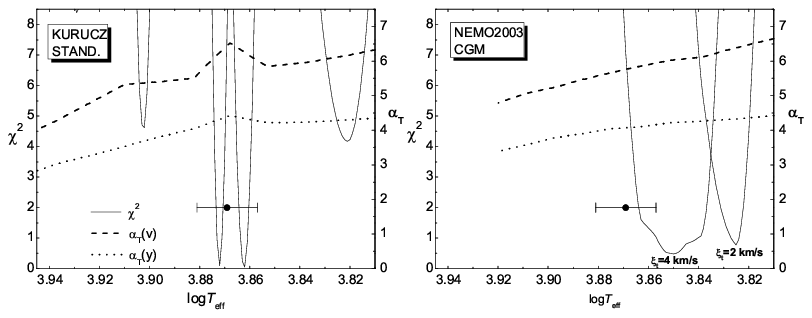}{ $\chi^2$ as a function of effective
temperature for a dominant mode of FG Vir derived using Kurucz
(left) and Vienna (right) models. Dot with error bar shows $\log
T_{\rm eff}$ derived from mean colours. The right linked layers
contain the temperature flux derivatives, $\alpha_T$, in $vy$
Str\"omgren passbands. In the right panel the effect of
microturbulence velocity is also shown.}
{fig3}{!ht}{clip,width=\textwidth}
\figureDSSN{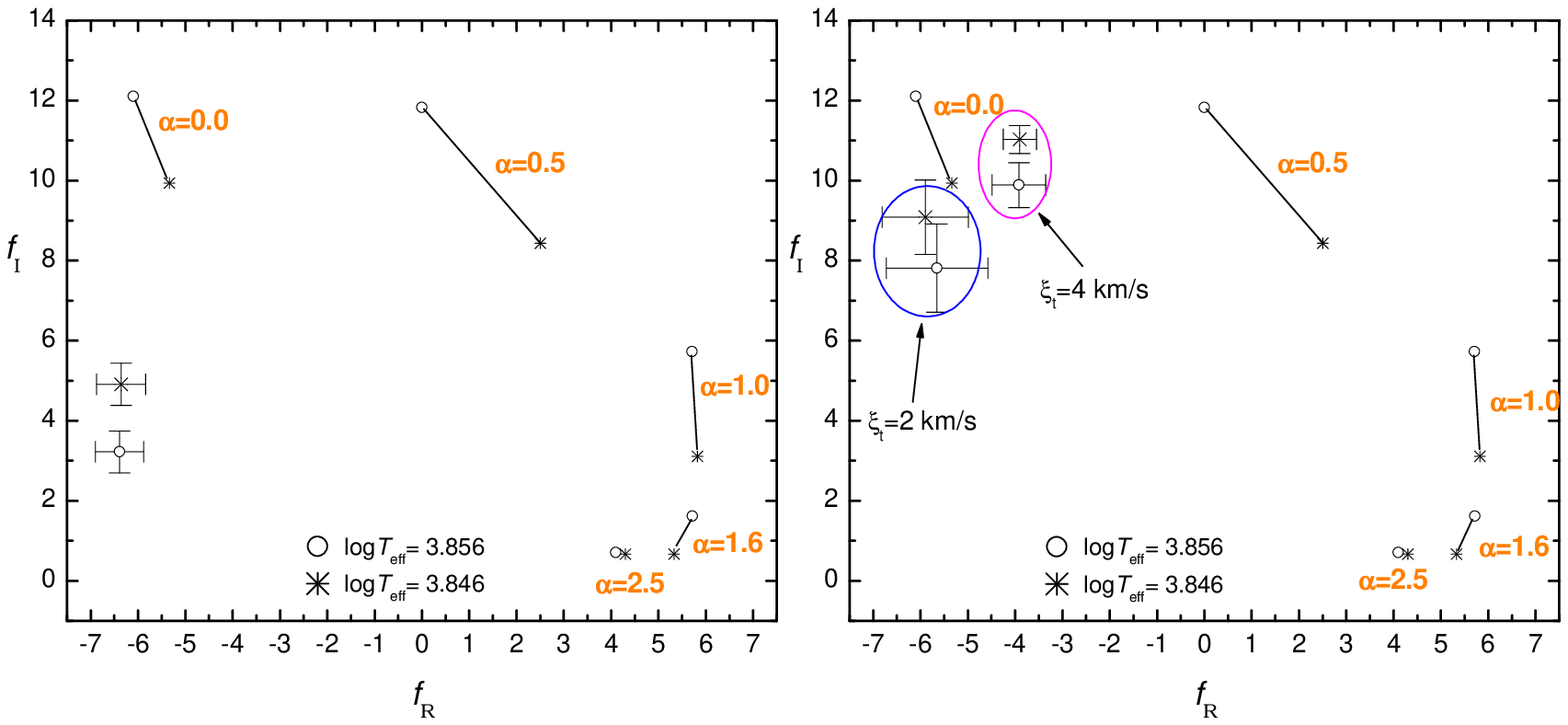}{Comparison of the $f$ values
inferred from Str\"omgren photometry for $\beta$ Cas with the
theoretical ones calculated with various MLT parameter, $\alpha$.
The empirical $f$ values were obtained adopting Kurucz models (left
panel) and Vienna models (right panel). In the right panel the
effect of microturbulence velocity is also shown.}
{fig4}{!ht}{clip,width=\textwidth}

\noindent
\section{Rotational mode coupling}
The most important effect of moderate rotation is mode coupling
(Soufi, Goupil, Dziembowski 1998). It takes place if the frequency
difference between modes $j$ and $k$ is of the order of angular
velocity of rotation, and if the spherical harmonic indexes satisfy
relations: $\ell_j=\ell_k+2$ and $m_j=m_k$. As eigenfunctions for
individual modes, we have to consider superpositions of all modes
satisfying the above conditions. Hence, the photometric amplitude of
a coupled mode is given by (Daszy\'nska-Daszkiewicz et al. 2002)
$${\mathcal A}_\lambda(i)= \sum_k a_k A_{\lambda,k}(i),$$
where the contribution from modes in non-rotating star is determined
by coefficients $a_k$ which are solutions from perturbation theory.
Now, the location of the mode on the diagnostic diagrams depends on
the azimuthal order, $m$, the inclination angle and the rotational
velocity. We considered the stellar model with the following
parameters $M=1.8M_{\odot}$, $\log T_{\rm eff}=3.866$ and $\log
L/L_\odot=1.12$, and the rotational velocity of about 70 km/s, which
are appropriate for FG Vir. As an example, we consider the
rotational coupling between $\ell=0$ and $\ell=2$ axisymmetric modes
with frequencies 19.342 and 19.597 d$^{-1}$, respectively. In Fig. 5
we show the position for coupled modes on the diagram with the
Str\"omgren $y$ passband and the radial velocity. The left panel
refers to the solution dominated by the $\ell=0$ component, whereas
the right one to the solution dominated by $\ell=2$. For discussion
of other effects of rotation within the perturbative approach see
e.g. Pamyatnykh (2003). \figureDSSN{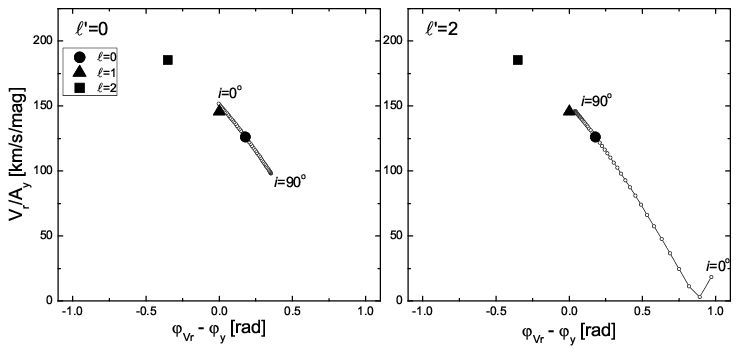}{ The
positions of rotational coupled modes (open circles) on the
$A_{Vr}/A_y~vs.~\varphi_{Vr}-\varphi_y$ diagram. We considered a
coupling between close $\ell=0$ and 2 pair at the rotation velocity
of about 70 km/s in the stellar model with $M=1.8M_{\odot}$ and
$\log T_{\rm eff}=3.866$. Filled symbols indicate positions of pure
= 0, 1, 2 modes.} {fig5}{!ht}{clip,width=\textwidth}

\noindent
\section{Summary}
I outlined results obtained in Paper I and II, where we proposed and
applied the new method of simultaneous determination of the
spherical harmonic degree, $\ell$, and the nonadiabatic parameter
$f$ from multi-colour photometry and radial velocity data. We
demonstrated that inferring $f$ values from such observations is
possible, thus identification of $\ell$ can be done without {\it a
priori} knowledge of $f$. Our method combines the photometry and
spectroscopy, and it gives the $\ell$ identification at the highest
confidence level achieved up to now. Moreover, by comparing
empirical and theoretical $f$ values, the method yields constraints
on mean stellar parameters and on properties of subphotospheric
layers. In the case of $\delta$ Sct stars, this is the treatment of
convective transport. Inferred values of $f$ are consistent with
models calculated assuming rather inefficient convection ($\alpha
\le 0.5$). The $f$ parameter constitutes a new asteroseismic tool
which is complementary to oscillation frequencies.

It is obvious that detecting more and more oscillation frequencies
is of great importance, especially in the satellite missions time.
However, it seems that asteroseismology of $\delta$ Sct stars will
be served better if we focus also on those frequencies for which
very accurate and simultaneous ground-based data from photometry and
spectroscopy can be obtained.

\vspace{10mm}

\acknowledgments{ The author thanks Wojtek Dziembowski and Alosha
Pamyatnykh for instructive comments and Miko{\l}aj Jerzykiewicz for
carefully reading the manuscript. This work was supported by the
Polish MNiI grant No. 1 P03D 021 28 and by the HELAS EU Network No.
026138.}

\References{
\rfr Balona L.\ A., Evers E.\ A., 1999, MNRAS, 302, 349
\rfr Breger M., Pamyatnykh A.\ A., Pikall H., Garrido R., 1999, A\&A, 341, 151
\rfr Breger M., Pamyatnykh A.\ A., Zima W., et al., 2002, MNRAS, 336, 249
\rfr Breger M., Lenz P., Antoci V., et al., 2005, A\&A, 435, 955
\rfr Daszy\'nska-Daszkiewicz J., Dziembowski W.\ A., Pamyantykh A.\ A., Goupil M.-J., 2002, A\&A, 392, 151
\rfr Daszy\'nska-Daszkiewicz J., Dziembowski W.\ A., Pamyatnykh A.\ A., 2003, A\&A, 407, 999 (Paper\,I)
\rfr Daszy\'nska-Daszkiewicz J., Dziembowski W.\ A., Pamyatnykh A.\ A., et al., 2005, A\&A, 438, 653 (Paper\,II)
\rfr Grigahc\`ene A., Dupret M.-A., Gabriel M., Garrido R., Scuflaire R., 2005, A\&A, 434, 1055
\rfr Dupret M.-A., Grigahc\`ene A., Garrido R., Gabriel M., Scuflaire R., 2005a, A\&A, 435, 927
\rfr Dupret M.-A., Grigahc\`ene A., Garrido R., et al., 2005b, MNRAS, 361, 476
\rfr Gough D.\ O., 1977, ApJ, 214, 196
\rfr Guzik J.\ A., Bradley P.\ A., 1995, Baltic Astron., 4, 482
\rfr Pamyatnykh A.\ A., Dziembowski W.\ A., Handler G., Pikall H., 1998, A\&A, 333, 141
\rfr Pamyatnykh A.\ A., 2003, Ap\&SS, 284, 97
\rfr Soufi F., Goupil M.-J., Dziembowski W.\ A., 1998, A\&A, 334, 911
\rfr Templeton M.\ R., Basu S., Demarque P., 2001, ApJ, 563, 999
\rfr Viskum M., Kjeldsen H., Bedding T.\ R., et al., 1998, A\&A, 335, 549
\rfr Zima W., Wright D., Bentley J., et al., 2006, A\&A, 455, 235
}

\end{document}